\documentclass{aastex}
\usepackage{epsfig}
\usepackage{emulateapj5}

\newcommand{\bp}{$\beta$~Pictoris}
\newcommand{\CI}{\ion{C}{1}}
\newcommand{\CIP}{\CI($^{3}$P)}
\newcommand{\CID}{\CI($^{1}$D)}
\newcommand{\NP}{$N(^{3}\rm{P})$}

\slugcomment{To appear in ApJ, 2000 August 1}

\shorttitle{CO and \CI\ in the \bp\ CS disk}
\shortauthors{Roberge et al.}

\begin{document}

\title{High resolution HST STIS spectra of \CI\ and CO in the \\
\objectname[HD39060]{$\beta$~Pictoris} circumstellar disk}

\author{A. Roberge and P. D. Feldman}
\affil{Department of Physics and Astronomy, Johns Hopkins University,
Baltimore, \\
Maryland 21218}
\email{akir@pha.jhu.edu}

\author{A. M. Lagrange}
\affil{Groupe d'Astrophysique de Grenoble, CERMO BP53X,
F-38041 Grenoble Cedex, France}
\email{Anne-Marie.Lagrange@obs.ujf-grenoble.fr}

\author{A. Vidal-Madjar and R. Ferlet}
\affil{Institut d'Astrophysique de Paris, CNRS, 98bis Bd Arago, 
F-75014 Paris, France}
\email{alfred@iap.fr}

\and

\author{A. Jolly\altaffilmark{1}, J. L. Lemaire, and F. Rostas}
\affil{DAMAP et URA 812 du CNRS, Observatoire de Paris-Meudon, F-92195
Meudon Cedex, France}
\email{jolly@lisa.univ-paris12.fr}

\altaffiltext{1}{Present Address: LISA, Universit\'{e} Paris 12, 94010 Creteil 
Cedex, 
France}

\begin{abstract}

High resolution FUV echelle spectra showing absorption features arising from 
\CI\ and CO gas in the \bp\ circumstellar (CS) disk were obtained on 1997 
December 6 and 19 using the Space Telescope Imaging Spectrograph (STIS). An 
unsaturated spin-forbidden line of \CI\ at 1613.376 \AA\ not previously seen 
in spectra of \bp\ was detected, allowing for an improved determination of the 
column density of \CI\ at zero velocity relative to the star (the stable 
component), 
\NP$~=~(2 - 4) \times 10^{16}$ cm$^{-2}$. Variable components with multiple 
velocities, which are the signatures of infalling bodies in the \bp\ CS disk, 
are 
observed in the \CI\ $\lambda\lambda$1561 and 1657 multiplets. Also seen for the 
first 
time were two lines arising from the metastable $^{1}$D level of carbon, at 1931 
\AA\ 
and 1463 \AA. The results of analysis of the CO $A-X$ (0-0), (1-0), and (2-0) 
bands 
are presented, including the bands arising from $^{13}$CO, with much better 
precision 
than has previously been possible due to the very high resolution provided by 
the STIS 
echelle gratings. Only stable CO gas is observed, with a column density 
$N(\rm{CO})~=~(6.3 \pm 0.3) \times 10^{14} \ \rm{cm}^{-2}$. An unusual ratio of 
the column 
densities of $^{12}$CO to $^{13}$CO is found ($R$~=~$15 \pm 2$). The large 
difference 
between the column densities of \CI\ and CO indicates that photodissociation of 
CO is not 
the primary source of \CI\ gas in the disk, contrary to previous suggestion.

\end{abstract}


\section{Introduction}

\bp\ is the most extensively studied of the young planetary systems discovered
in the last decade and a half. It is a bright Southern hemisphere star 
(type A5 V), located about 19.3 pc distant from the Sun, with a
systemic radial velocity of 20 km s$^{-1}$ (for a review of the \bp\ system, see 
\citet{vid98}). It was observed in 1983 by the IRAS satellite to have a 
large excess of emission at infrared wavelengths.  This was referred to as 
the Vega-like phenomenon and was identified as arising from an edge-on 
circumstellar (hereafter CS) dust disk, presumed to be associated with
planetary formation \citep{smi84}. It was soon determined through absorption 
spectroscopy that there was CS gas associated with the dust as well.

A large body of evidence has accumulated indicating that there are 
comet-like bodies present in the \bp\ CS disk.  Collisions between dust 
particles are expected to produce submicron fragments which should be expelled 
from the system by radiation pressure on time scales much shorter than any 
plausible stellar age. Thus there must be a secondary source of particles; one 
model for the production of dust and gas in the CS disk focusses on evaporating 
comets and is called the Orbiting Evaporating Bodies model (OEB). In this 
picture, the comets orbit the star at several tens of AU, and thus, the \bp\ CS 
disk is a kind of ``gigantic multi-cometary tail with its natural constituents: 
gas and dust'' \citep{lec96}. Spectra of \bp\ show variable 
redshifted absorption features arising from gas infalling toward the star at 
high velocities (and infrequently, blueshifted features as well); these features 
are best attributed to the evaporation of star-grazing comets, called the 
Falling 
Evaporating Bodies scenario (FEB) \citep{beu90}. Also, gas at close to 20 km 
s$^{-1}$ (zero radial velocity relative to the star) is identified in all 
observations and is called the stable gas component. This gas is 
difficult to understand, as modeling indicates that it should be expelled from 
the 
system by radiation pressure; a continuous source for this gas is required.

Neutral carbon and carbon monoxide have been observed in HST-GHRS UV absorption
spectra of \bp; carbon monoxide is the only molecule detected in the CS disk to 
date \citep{vid94, jol98}. Since CO and \CI\ can be dissociated and ionized by 
interstellar UV photons on time scales of the order of 200 years, both must
be continuously replenished. Carbon monoxide, in particular, is difficult to 
reform after dissociation in the \bp\ environment. Thus, the presence of these 
species indicates that a secondary source for this gas should exist, just as for 
the CS dust \citep{vid94}. \citet{jol98} found the column densities of \CI\ and 
CO to be comparable, around $10^{15}$ cm$^{-2}$; since their rates of 
destruction are also comparable, this was taken as evidence that the \CI\ is 
produced by photodissociation of CO, which evaporates from comets orbiting at 
various distances and velocities. However, the \CI\ column density was 
determined from a heavily saturated multiplet and is therefore quite uncertain. 
In the hopes of further constraining the characteristics of \CI\ and CO in the 
\bp\ system, we have reinvestigated the transitions observed in the GHRS data, 
as well as some that were not seen due to the relatively low spectral resolution 
of GHRS compared with that of the STIS high resolution echelle.

\section{Observations}

HST STIS high resolution echelle spectra of \bp\ were obtained on 1997 
December 6 and 1997 December 19, covering the wavelength range from 1459 \AA\ 
to 2888 \AA\ in six exposures each day. Table~1 shows the log of observations.
All the absorption features discussed in this paper appear in either the first 
or second data set listed for each day (o4g001010/o4g002010 or 
o4g001020/o4g002020).The data were initially reduced and calibrated using the 
STScI IRAF package \emph{calstis v1.8}.  Spectra with a signal-to-noise ratio of 
around 10 were achieved.  Examination of the errors in the flux values showed 
that the error propagation calculation had not been performed correctly and that 
thestated errors were too small. Therefore the data were re-calibrated using 
\emph{calstis v1.9a}, correcting the underestimate of the measurement errors.

The particular advantage of this data set over previous comparable ones is
the very high spectral resolution achieved. The instrumental line spread 
function using the E140H grating is well described by a Gaussian with a 
FWHM of 1 pixel, corresponding to FWHM~=~$\lambda$/220,000 \citep{sah99}. 
However, since the detector undersamples the line spread function, the 
effective resolution to separate two adjacent lines with this 
grating is only $R$~=~110,000. Using the E230H grating, the FWHM is about 2 
pixels, corresponding to FWHM~=~$\lambda$/110,000 \citep{sah99}.

\section{Analysis}

\subsection{\CI] $\lambda$1613 line}

This previously undetected spin-forbidden transition $(^{3}$P$_{0}\: 
\rightarrow \:^{1}$P$_{1})$ at 1613.376 \AA\ was clearly observed on both days. 
It is unsaturated, which allows for an improved determination of the column 
density of \CI. As there was no change in the line between the two days of 
observation, the spectra were averaged together to improve the S/N; the result 
is shown in Figure~1. Unfortunately, the other two fine structure lines in the 
multiplet were not reliably detected. Atomic data were taken from \citet{mor91} 
and the continuum around the line was fit with a sixth-degree polynomial. Voigt 
line profiles were used to generate transmission functions, which were then 
convolved with a Gaussian instrumental line spread function with 
FWHM~=~$\lambda$/220,000 to create model spectra. The $\chi^{2}$ statistic 
between the model and the data was then minimized to determine the best velocity 
centroid, $v$, column density of \CI\ in the ground level, $N(^{3}\rm{P}_{0})$, 
and Doppler broadening parameter, $b$; error bars for these parameters were 
determined from the contours of $\chi^{2}$. Since the line was unsaturated, 
the column density was also determined from the equivalent width of the line 
and $b$ and $v$ were determined from a simple Gaussian fit to the line, as a 
check 
on the results of the $\chi^{2}$ minimization.

\subsection{\CIP\ $\lambda\lambda$1561 and 1657 multiplets}

The central portion of the 1561 \AA\ and 1657 \AA\ multiplets, arising from 
the $^{3}$P ground term, were heavily saturated in the STIS data. However, this 
allowed us to examine the smaller variable red and blueshifted features in the 
spectra which are the signatures of infalling comets in the CS disk. In 
Figures~2 
and 3, one can easily see absorption lying at higher redshift in the December 6 
data that is not visible in the December 19 data. A model of the multiplet which 
contains only one velocity component is unable to reproduce the absorption dip 
at 1561.5 \AA\ or the absorption at higher redshift in the December 6 data.

Due to the fact that the multiplets were extremely saturated and that the 
multiple velocity components were blended together, we were not able to perform 
$\chi^{2}$ minimizations to find unique best values for the model parameters. 
\ion{Fe}{2} absorption features in our data set were analyzed and the velocities 
of 
their multiple velocity components found. Models of the \CI\ $\lambda$1561 
multiplet 
were constructed, containing the same number of components as were found in the 
\ion{Fe}{2} features, with roughly the same velocities. The parameters of 
the stable component at $v$ = 20 km s$^{-1}$ were set to those determined from 
the 
analysis of the unsaturated 1613.376 \AA\ line. The remaining parameters were 
then 
adjusted to find the best model by eye. This model was compared to the 1657 \AA\ 
multiplet to confirm that the values found in this way were reasonable.  A small 
constant value ($\approx 2 \times 10^{-13}$ erg s$^{-1}$ cm$^{-2}$ \AA$^{-1}$) 
was subtracted from the 1657 \AA\ data to bring the baselines of the totally 
saturated features to zero.

\subsection{\CID\ $\lambda\lambda$1931 and 1463 lines}

These lines arise from the excited, metastable $^{1}$D state of the ground 
configuration and have not previously been observed in spectra of \bp. The 
$^{1}$D level has a lifetime of about 4000 s, lying about 10,000 cm$^{-1}$ 
above the ground level; C atoms in this state may be produced during the 
photodissociation of CO in solar system comet comae \citep{toz98}. The 
1931 \AA\ lines were analyzed using the same model generating and $\chi^{2}$ 
minimization procedures described in \S\ 3.1; the atomic data used were from 
\citet{hib93}. Although there may be some reason to suspect that the strong 
central absorption features near 1931.05 \AA\ contain multiple velocity 
components, since this behavior is seen in lines arising from excited levels 
of other atomic species, the $\chi^{2}$ minimizations indicated that stable 
unique solutions containing more than one component in the central absorption 
features could not be found. Thus, best models were found using only one 
velocity 
component. However, the difficulty involved in modeling a saturated, blended 
line 
is such that there may well be other undetected components present, so models 
with two components in the central absorption feature were compared to the data 
by eye. The 1463 \AA\ spectra, shown in Figure~5, were so noisy that they served 
only to roughly confirm the values found from the 1931 \AA\ line.  A small 
constant 
flux value ($\approx 1 \times 10^{-13}$ erg s$^{-1}$ cm$^{-2}$ \AA$^{-1}$) was 
subtracted from the 1463 \AA\ data to bring the baselines of the saturated lines 
to zero. 

\subsection{CO Fourth Positive band system}

The (0-0), (1-0), and (2-0) bands of the Fourth Positive system of CO 
($A \: ^{1}\Pi \:-\: X \:^{1}\Sigma ^{+}$) did not vary between the two days of 
observation. The spectra were therefore averaged together to improve the S/N; 
the 
result appears in Figure~6. Only a single velocity component was observed. 
Note the detection of bands arising from $^{13}$CO, the perturbation band 
$e \: ^{3}\Sigma^{-} \:-\: X \:^{1}\Sigma ^{+}$ (1-0), not 
previously observed in spectra of \bp, and the strong perturbation band 
$d \: ^{3}\Delta \:-\: X \:^{1}\Sigma ^{+}$ (5-0).

Models were generated using wavelengths and oscillator strengths from 
\citet{mor94}. Energies of the ground state levels were calculated using the 
Dunham coefficients from \citet{far91} and LTE assumed in order to 
determine the population of the rotational levels. The parameters of the model 
are the rotational excitation temperature, $T$, the Doppler broadening
parameter, $b$, the column density of $^{12}$CO in the ground vibrational state, 
$N(^{12}\rm{CO})$, the column density of $^{13}$CO in the ground vibrational 
state, 
$N(^{13}\rm{CO})$, and the velocity centroid, $v$. $\chi^{2}$~minimization was 
then 
performed on all three bands simultaneously.

\section{Results}

\subsection{\CIP\ stable component}

The \CI] $\lambda$1613.376 line showed only one velocity component at 20 km 
s$^{-1}$, the 
systemic velocity of the star; the nominal uncertainty in the absolute 
wavelength 
calibration of STIS leads to an error in velocity determinations of about 
1 km s$^{-1}$ \citep{sah99}. This velocity and the fact that the line did not 
change 
between the two days of observation identifies the line as arising from 
stable gas. The results of the $\chi^{2}$ minimizations for all the features 
analyzed appear in Table~2. The column density for stable carbon in the ground 
level, $^{3}$P$_{0}$, determined from $\chi^{2}$~minimization was found to agree  
with that determined from the equivalent width of the line. Similarly the $v$ 
and $b$ 
determined from $\chi^{2}$ minimization agreed with the values found from 
fitting 
a simple Gaussian to the unsaturated line.  

Using a 3-$\sigma$ upper limit on the column density of \CI\ in the 
$^{3}$P$_{2}$ 
level (from the non-detection of the fine structure line at 1614.5068 \AA) and 
assuming LTE, a firm upper limit of 100 K on the excitation temperature of 
$^{3}$P carbon in the stable component was found. From analysis of the 1561 \AA\ 
multiplet, discussed below, it was found that the excitation temperature of the 
stable component must be greater than about 50 K, or the multiplet could not be 
reasonably modeled. This range in temperature allows us to determine that the 
total column density of stable \CI\ in the $^{3}$P ground term is 
$(2~-~4)~\times~10^{16}$~cm$^{-2}$.

This column density is more than an order of magnitude larger than the total 
$^{3}$P column density found by \citet{jol98} from GHRS data taken in November 
1994, \NP$~=~2 \times 10^{15}$ cm$^{-2}$. Either the abundance of stable 
\CI\ has varied over the three years between observations or, as is more likely, 
the difficult modeling of the heavily saturated \CI\ $\lambda$1561 multiplet in 
the GHRS data led to an inaccurate column density.  Models with column densities 
of $10^{16}$ cm$^{-2}$ in the stable component were compared to the 1994 GHRS 
spectrum of the 1561 \AA\ multiplet and found to fit the data equally well as 
models with the lower column density. We found the Doppler broadening parameter 
of the stable \CIP\ to be $1.3 \pm 0.5 \ \rm{km} \ \rm{s}^{-1}$.  The 
previous work on the GHRS data found a Doppler broadening parameter of 4.2 
km s$^{-1}$, which was much greater than the parameter found for the other 
atomic species and CO \citep{jol98}. Our smaller $b$ value is equal to that 
found for 
CO in our data; we thus do not see any excess kinetic energy in the motions of 
the 
\CI\ atoms. 

\subsection{\CIP\ multiple velocity components}

The difficulty in modeling a saturated, blended multiplet is formidable, and the 
values determined for the multiple velocity components of \CIP\ from the 
1561 \AA\ multiplet are not reliable. However, this analysis did confirm that 
models with column densities of $10^{16}$ cm$^{-2}$ in the stable component 
could 
reasonably fit the 1561 \AA\ spectra. We can also conclude that the total column 
density of \CIP\ in the variable components is about $10^{14}$ cm$^{-2}$ on 
December 6 and about $10^{15}$ cm$^{-2}$ on December 19, to within 
an order of magnitude or so. It is also clear that the variable \CI\ features 
are 
generally better fit with higher excitation temperatures and larger $b$ 
parameters, 
but these values are very poorly determined. 

\subsection{\CID}

Using models with only one velocity component in the strong central absorption 
features, two components arising from the excited $^{1}$D 
level of \CI\ were found in the December 6 data, only one in the December 19 
data.  
The best models are overplotted on the spectra in Figures~4 and 5.  Note that 
the 
central absorption features do not appear at the same velocity on the two days, 
that they 
also do not appear at the systemic velocity of the star, and that the column 
density of \CID\ changes significantly between the two days of observation. 
This would seem to indicate that there is no stable component for \CID. 
Models with two velocity components in the central absorption feature, which 
were 
compared to the data by eye, indicate that there could be a ``stable'' 
component which has the same column density on both days. However, it must be at 
about 22-23 km s$^{-1}$, which is significantly different from the systemic 
velocity 
of \bp. These data are not able to conclusively determine the velocity structure 
present in the \CID\ gas; observation of unsaturated lines at similarly high 
resolution will be necessary.

\subsection{CO}

The CO bands showed no multiple velocity component structure and no change 
between the two days of observation; the absorbing gas is thus entirely 
associated with the stable component at the systemic velocity of the star. The 
best model is shown overplotted on the data in Figure~6. The rotational 
excitation temperature and the ratio of $N(^{12}\rm{CO})$ to $N(^{13}\rm{CO})$ 
show no significant change from the values found by \citet{jol98}. 
$R(^{12}\rm{CO}/^{13}\rm{CO})$~=~$15 \pm 2$ is quite small compared to typical 
values found in the ISM, e.g. $R$~=~$150 \pm 27$ for the diffuse clouds toward 
$\zeta$~Ophiuchi \citep{she92}. However, chemical fractionation at low kinetic 
temperatures can explain this unusual ratio, as discussed below. The column 
densities of both $^{12}$CO and $^{13}$CO were found to be about 1/3 the values 
found from the GHRS data. Note that the column density of $^{12}$CO is about 50 
times smaller than the column density of stable \CIP.

\section{Discussion}

Although CO in solar system comet comae is photodissociated by solar FUV 
photons, the CO in the \bp\ CS disk is primarily destroyed by interstellar 
photons. The dissociation energy for CO is 11.1 eV; thus CO may only be 
dissociated by photons with wavelengths shortward of $\sim$~1100 \AA. The type 
A5 star \bp\ lacks the strong FUV emission lines created in the Sun's 
chromosphere; thus it emits very little flux in the FUV. The only source for 
CO-dissociating photons at \bp\ is therefore the interstellar UV radiation 
field. Since CO photodissociates primarily through discrete line absorptions, 
self-shielding can have a strong effect on the abundance of CO in interstellar 
clouds \citep{van88}. However, this is unlikely to occur in the \bp\ CS disk 
because of the very small transverse dimension of the disk.  Thus, the 
photodissociation rate for $^{12}$CO in the \bp\ disk should be equal to the 
unshielded value and should also be equal to the value for $^{13}$CO. Since 
$^{13}$CO is not selectively dissociated in this situation, its high abundance 
was explained by the reaction \[ ^{13}\rm{C}^{+} + ^{12}\rm{CO} \;
\rightleftharpoons \; ^{13}\rm{CO} + ^{12}\rm{C}^{+} + 35 \ \rm{K} \] 
which favors the production of $^{13}$CO at gas kinetic temperatures below 35 K 
\citep{jol98}. Assuming chemical equilibrium and that isotopic exchange is much 
more important than photodissociation for both $^{12}$CO and $^{13}$CO 
\citep{she92}, 
\[ \frac{n(^{12}\rm{CO})}{n(^{13}\rm{CO})} =  \exp 
\left(\frac{-35}{T_{kin}}\right) \left(\frac{^{12}\rm{C}}{^{13}\rm{C}}\right) = 
15 \pm 2. \] The rotational excitation temperature of CO often does not 
accurately describe the gas kinetic temperature in diffuse environments 
\citep{wan97}. However, assuming a $(^{12}\rm{C}/^{13}\rm{C})$ ratio, we may use 
the above expression to estimate the true gas kinetic temperature of the CO.

The average carbon isotopic ratio in the local ISM is $\sim \ 60-70$ 
\citep{lan93} and the typical solar system value found in comets is 
$(^{12}\rm{C}/^{13}\rm{C}) \ = \ 89$ \citep{jew97}. Using the range 
$(^{12}\rm{C}/^{13}\rm{C}) \ = \ 89-60$, we find that the gas kinetic 
temperature of the carbon monoxide is 20~K $ - $ 25~K, indicating that the CO 
gas is indeed colder than the stable \CI\ gas. This suggests that the \CI\ and 
CO are not located in the same regions of the disk. The assumption that isotopic 
exchange is more important than photodissociation should be reasonable. For this 
assumption to apply, \[ \Gamma \; \ll \; k^{f} \exp 
\left(\frac{-35}{T_{kin}}\right) \ n(^{12}\rm{C}^{+}), \] where $\Gamma$ is the 
unshielded photodissociation rate, $2 \times 10^{-10} \ \rm{s}^{-1}$ 
\citep{van88}, k$^{f}$ is the forward reaction rate, $6.8 \times 10^{-10}  \ 
\rm{cm}^{3} \ \rm{s}^{-1}$ at 80 K \citep{smi80}, and n$(^{12}\rm{C}^{+})$ is 
the volume density of $^{12}\rm{C}^{+}$. The average volume density of C, 
$n(^{12}\rm{C}) \ = \ \it{N}(\rm{^{3}P})/\it{r} \ \simeq \ \rm{20 \ cm}^{-3}$, 
assuming that the carbon extends over a distance $r$ = 100 AU. Further assuming 
that $n(^{12}\rm{C}^{+}) \geq \it{n}(\rm{^{12}C})$, as is likely, the right hand 
side of the above expression is greater than or equal to about $2 \times 10^{-9} 
\ \rm{s}^{-1}$, an order of magnitude larger than $\Gamma$. However, a more 
complete treatment of the relationship between $(^{12}\rm{CO}/^{13}\rm{CO})$ and 
$(^{12}\rm{C}/^{13}\rm{C})$ takes into account photodissociation but requires 
exact knowledge of the density of $^{12}\rm{C}^{+}$ in the CS disk 
\citep{she92}.

The much larger column density of \CIP\ compared to that of CO leads us to 
believe that photodissociation of CO cannot be the only source of stable \CI\ in 
the \bp\ CS disk. Since there is no evidence that CO is produced by infalling 
comets (no variability or red and blueshifted features), it has been postulated 
that the CO gas slowly evaporates from the OEBs at several tens of AU from the 
star and is photodissociated to produce the stable 
\CI\ \citep{lec98}. Obviously some portion of the \CI\ gas must be produced 
directly from the FEBs (the portion giving rise to the variable red and 
blueshifted absorption); perhaps this \CI\ gas is decelerated somehow and 
accumulates in the stable component before being destroyed by photoionization. 
In this scenario, the equilibrium column density of \CI\ is \[ N(^{3}\rm{P}) = 
\frac{\it{n} \times N_{FEB}}{\Gamma} \] where $n$ is the mean number of 
infalling comets per year, $\sim 10^{2}$ per year \citep{vid98}, $N_{FEB}$ is 
the total column density of \CI\ gas produced by an infalling comet, and 
$\Gamma$ is the photoionization rate for \CI, 0.004 yr$^{-1}$. This expression, 
which assumes that \CI\ atoms are lost only through photoionization by 
interstellar UV photons, gives $N_{FEB} \sim 10^{11}$ cm$^{-2}$, which should 
easily be produced by infalling comets like the ones giving rise to the variable 
\CIP\ components in our data. However, this is a rough treatment and the number 
of infalling comets per year varies significantly over time scales of a few 
years \citep{vid98}. Also, this treatment does not take into account reformation 
of \CI\ by radiative recombination; we cannot take this process into account 
properly without knowledge of the amount of \ion{C}{2} in the CS disk and the 
electron density.

Considering the \CID, when such atoms are produced by photodissociation of CO, 
\ion{O}{1}($^{1}$D) atoms must be produced also to conserve spin. The minimum 
total photon energy needed for this dissociation is 14.33 eV, corresponding to a 
threshold wavelength of 865 \AA. Since this threshold is below the Lyman limit, 
there are virtually no interstellar UV photons capable of producing \CID\ by 
photodissociation of CO. Thus, the \CID\ atoms in the \bp\ CS disk cannot be 
produced by photodissociation of CO by stellar or interstellar photons and must 
be produced by a collisional process involving ground state carbon atoms. Since 
the energy of the $^{1}$D state relative to the ground state is high, the 
collisional process must be a very energetic one and therefore is likely to be 
closely associated with the infalling bodies. Consequently, with a short \CID\ 
lifetime, there would be no stable component in this gas; this behavior has not 
been previously seen in any constituent of the \bp\ CS disk. But this result is 
tentavive; examination of an unsaturated line at high resolution is needed to 
determine the velocity structure of the \CID\ gas.

\section{Concluding Remarks}

The very high resolution, low scattered light contamination, and good order 
separation of this STIS echelle data set has provided some clear advantages over 
previous observations of \bp. The rotational lines of CO have been resolved, 
allowing for a much more precise determination of the physical parameters of the 
gas. The column density of CO is $N(\rm{CO})~=~(6.3 \pm 0.3) \times 10^{14} 
\ \rm{cm}^{-2}$ and the ratio $R(^{12}\rm{CO}/^{13}\rm{CO})~=~15 \pm 2$ is 
found. The absence of transient red or blueshifted components in the high 
resolution CO spectra supports the suggestion that this gas evaporates from 
cometary bodies orbiting far (several tens of AU) from the star. But the fact 
that the column density of CO is only about 2\% of the total column density of 
\CI\ in the $^{3}$P ground term implies that photodissociation of this CO is not 
the primary source for \CI\ gas. It could perhaps be produced directly from 
infalling comets close to the star, but the mechanism by which it comes to zero 
velocity relative to the star and accumulates before being photoionized is 
unclear (although see \citet{lag98}). The \CID\ gas may not have a stable 
component at 20 km s$^{-1}$; this unique species could prove to be a valuable 
tracer of FEB activity in the \bp\ CS disk. 

Despite the advantages of this data set, our lack of success in modeling the 
heavily saturated \CI\ multiplets indicates that in order to really determine 
the characteristics of the variable components of the \CI\ gas, we need to 
observe an unsaturated line or multiplet, with an oscillator strength between 
that of the 1561 \AA\ multiplet and that of the spin-forbidden 1613.376 \AA\ 
line. A number of suitable multiplets and lines lie in the FUV, shortward of 
$\sim$ 1300 \AA.  Also, an unsaturated line arising from the $^{1}$D level would 
allow us to confirm the velocity structure in this gas and to determine if the 
velocities of the $^{1}$D gas components correspond with any of the velocities 
of the variable components in the $^{3}$P gas. Three likely lines lie between 
1311 \AA\ and 1359 \AA. Measurement of the densities of \ion{C}{2} and 
\ion{O}{1} would greatly help to unravel the carbon chemistry of the \bp\ disk. 
Again, potentially useful multiplets of these species lie in the FUV below 1340 
\AA. Thus, although this data set has vastly increased our knowledge about the 
important species \CI\ and CO in the \bp\ disk, our understanding would probably 
benefit greatly from investigation of \bp\ at shorter ultraviolet wavelengths.

\acknowledgements

We thank Jason McPhate for his work on our absorption line profile codes and 
John Debes for his work on the \ion{Fe}{2} lines. We also thank B-G Andersson 
and our reviewer, X. Tielens, for their fruitful comments. This work is based 
on observations with the National Aeronautics and Space Administration -- 
European Space Agency HST obtained at the Space Telescope Science Institute, 
which is operated by the Association of Universities for Research in Astronomy, 
Incorporated, under NASA contract NAS5-26555. Support for this work at JHU was 
provided by grant GO-07512.01-96A from the Space Telescope Science Institute.

\clearpage

\begin{deluxetable}{cccccc}
\tablecolumns{6}
\tablewidth{0pt}
\tablecaption{HST-STIS observation log}
\tablehead{\colhead{Observation} & \colhead{Grating} & \colhead{Slit} & 
\colhead{Spectral Range} & 
\colhead{Start Time} & \colhead{Exposure Time} \\
\colhead{ID \#} & \colhead{} & \colhead{(arcsec)} & \colhead{(\AA)} & 
\colhead{(UT)} & \colhead{(s)}}
\startdata
\sidehead{1997 Dec 6}
o4g001010 & E140H & $0.2 \times 0.09$ & 1461 -- 1663 & 
07:17:58 & 900.0 \\ 
o4g001020 & E230H & $0.1 \times 0.03$ & 1874 -- 2146 & 
07:40:26 & 80.0 \\ 
o4g001030 & E230H & $0.1 \times 0.09$ & 1624 -- 1896 &
07:48:20 & 678.8 \\ 
o4g001040 & E230H & $31 \times 0.05$ & 2624 -- 2895 &
08:48:28 & 360.0 \\ 
o4g001050 & E230H & $31 \times 0.05$ & 2374 -- 2646 &
09:02:20 & 360.0 \\ 
o4g001060 & E230H & $31 \times 0.05$ & 2124 -- 2396 &
09:16:12 & 288.0 \\
\tableline  
\sidehead{1997 Dec 19}
o4g002010 & E140H & $0.2 \times 0.09$ & 1461 -- 1663 &
19:51:29 & 900.0 \\ 
o4g002020 & E230H & $0.1 \times 0.03$ & 1874 -- 2146 &
20:13:57 & 80.0 \\ 
o4g002030 & E230H & $0.1 \times 0.09$ & 1624 -- 1896 &
20:21:51 & 678.8 \\ 
o4g002040 & E230H & $31 \times 0.05$ & 2624 -- 2895 &
21:25:45 & 360.0 \\ 
o4g002050 & E230H & $31 \times 0.05$ & 2374 -- 2646 &
21:39:37 & 360.0 \\ 
o4g002060 & E230H & $31 \times 0.05$ & 2124 -- 2396 &
21:53:29 & 288.0 \\  
\enddata 
\end{deluxetable}

\clearpage

\begin{deluxetable}{lcccc}
\tablecolumns{5}
\tablewidth{0pt}
\footnotesize
\tablecaption{Results of Modeling}
\tablehead{\colhead{} & \colhead{Velocity} & \colhead{Column density} & 
\colhead{Doppler broadening parameter} & \colhead{Temperature} \\
\colhead{} & \colhead{$v \ (\rm{km \ s}^{-1})$} & \colhead{$N \ (\rm{cm}^{-2})$} 
& 
\colhead{$b \ (\rm{km \ s}^{-1})$} & \colhead{$T (\rm{K})$}}
\startdata
\sidehead{Stable \CI($^{3}\rm{P}_{0}$)}
 & $20 \pm 1$ & $(6 \pm 1) \times 10^{15}$ & $1.3 \pm 0.5$ & \nodata \\
\tableline
\sidehead{\CIP}
December 6 & $20 \pm 1$ & $(2 - 4) \times 10^{16}$ & $1.3 \pm 0.5$ & 50 -- 100 
\\
           & 26 & $10^{14}$ & 4 & $10^{2}$ \\
           & 41 & $10^{14}$ & 10 & $10^{3}$ \\
           & 57 & $10^{13}$ & 12 & $10^{3}$ \\
 	   &    &           &    &          \\
December 19 & 11 & $10^{13}$ & 5 & $10^{1}$ \\
            & $20 \pm 1$ & $(2 - 4) \times 10^{16}$ & $1.3 \pm 0.5$ & 50 -- 100 
\\
            & 27 & $10^{15}$ & 5 & $10^{2}$ \\
\tableline
\sidehead{\CID}
December 6\tablenotemark{a} & $23 \pm 1$ & $(4 \pm 3) \times 10^{13}$ & $(4 \pm 
1)$ & 
\nodata \\
          & $49 \pm 2$ & $(4 \pm 2) \times 10^{12}$ & $(4 \pm 2)$ & \nodata \\
          & & & & \\
December 6\tablenotemark{b} & $22$ & $4 \times 10^{13}$ & $2$ & \nodata \\
          & $28$ & $8 \times 10^{12}$ & $3$ & \nodata \\
          & $48$ & $6 \times 10^{12}$ & $3$   & \nodata \\
          & & & & \\*
December 19\tablenotemark{a} & $26 \pm 1$ & $(1.2 \pm 0.7) \times 10^{14}$ & 
$4.4 \pm 0.8$ 
& \nodata \\
& & & & \\
December 19\tablenotemark{b} & $23$ & $4 \times 10^{13}$ & $2$ & \nodata \\
         & $29$ & $3 \times 10^{13}$ & $3$ & \nodata \\
\tableline
\sidehead{$^{12}$CO}
  & $20 \pm 1$ & $(6.3 \pm 0.3) \times 10^{14}$ & $1.3 \pm 0.1$ & $15.8 \pm 0.6$ 
\\
\tableline
\sidehead{$^{13}$CO}
 & $20 \pm 1$ & $(4.3 \pm 0.4) \times 10^{13}$ & $1.3 \pm 0.1$ & $15.8 \pm 0.6$ 
\\
\enddata
\tablenotetext{a}{Results from $\chi^{2}$ minimizations using model with one 
velocity 
component in the central absorption features}
\tablenotetext{b}{Results from chi-by-eye using model with two velocity 
components in the 
central absorption features.}
\end{deluxetable}

\clearpage
\newpage

\begin{figure}
\begin{center}
\epsfig{file=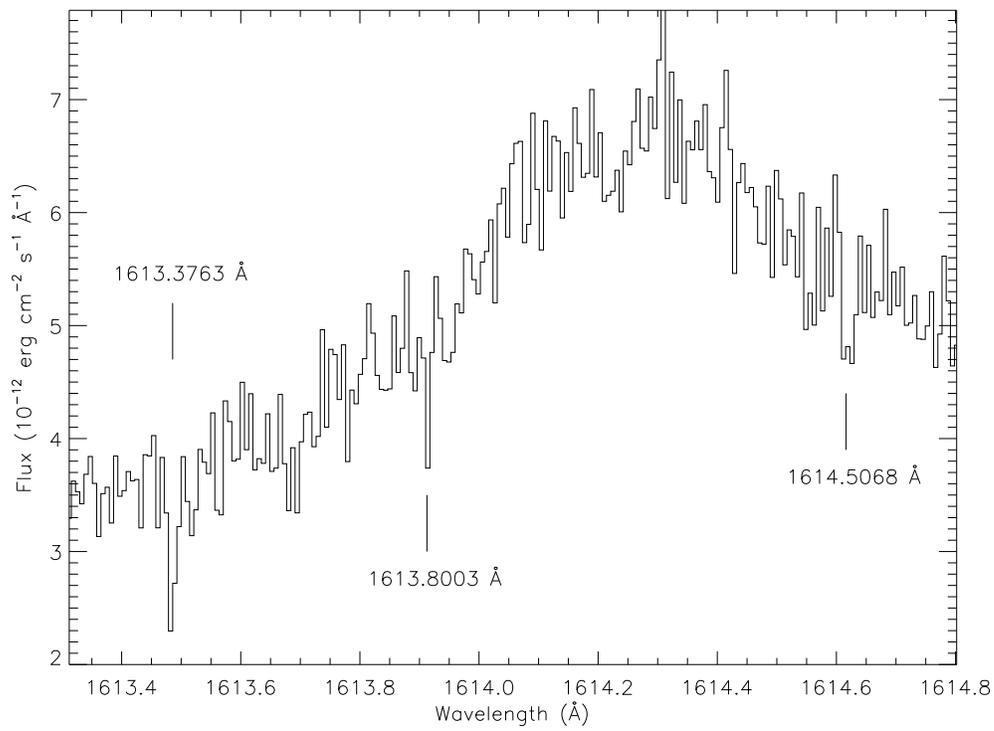, height=5.5in, angle=90}
\caption{The \CI] $\lambda$1613.376 line, with both days of data 
averaged together. Positions of the two other non-detected fine structure 
lines in the multiplet are indicated as well. The spectrum has not been rebinned 
or smoothed. \label{fig1}}
\end{center}
\end{figure}

\clearpage
\newpage

\begin{figure}
\begin{center}
\epsfig{file=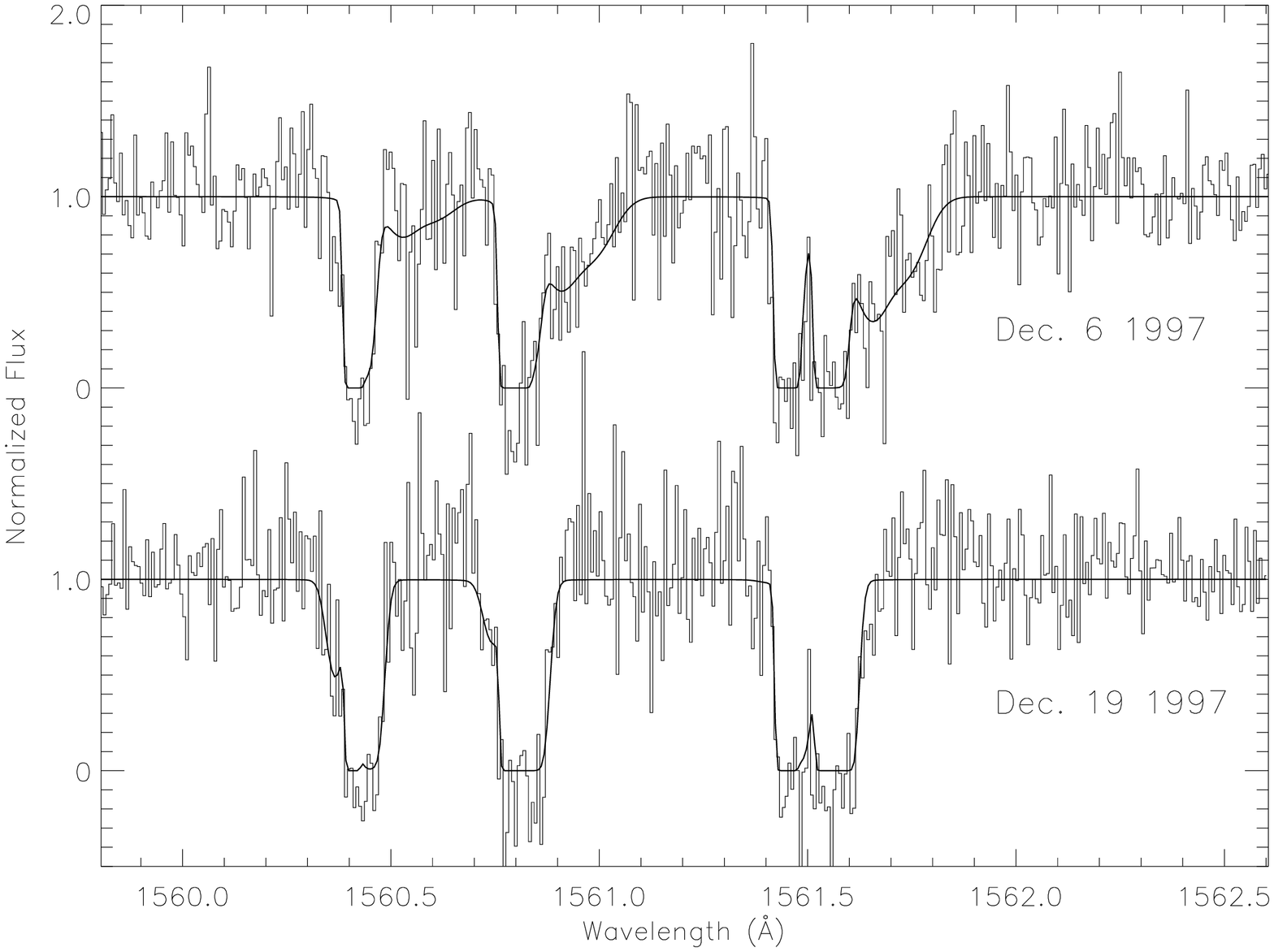, height=5.5in, angle=90}
\caption{The \CI\ $\lambda$1561 multiplet, with both days of data 
shown. The data have not been rebinned or smoothed. The continua around the 
features were fit with sixth-degree polynomials and the best fitting models based 
on the \CIP\ parameters in Table 2 are shown overplotted on the data. \label{fig2}}
\end{center}
\end{figure}

\clearpage
\newpage

\begin{figure}
\begin{center}
\epsfig{file=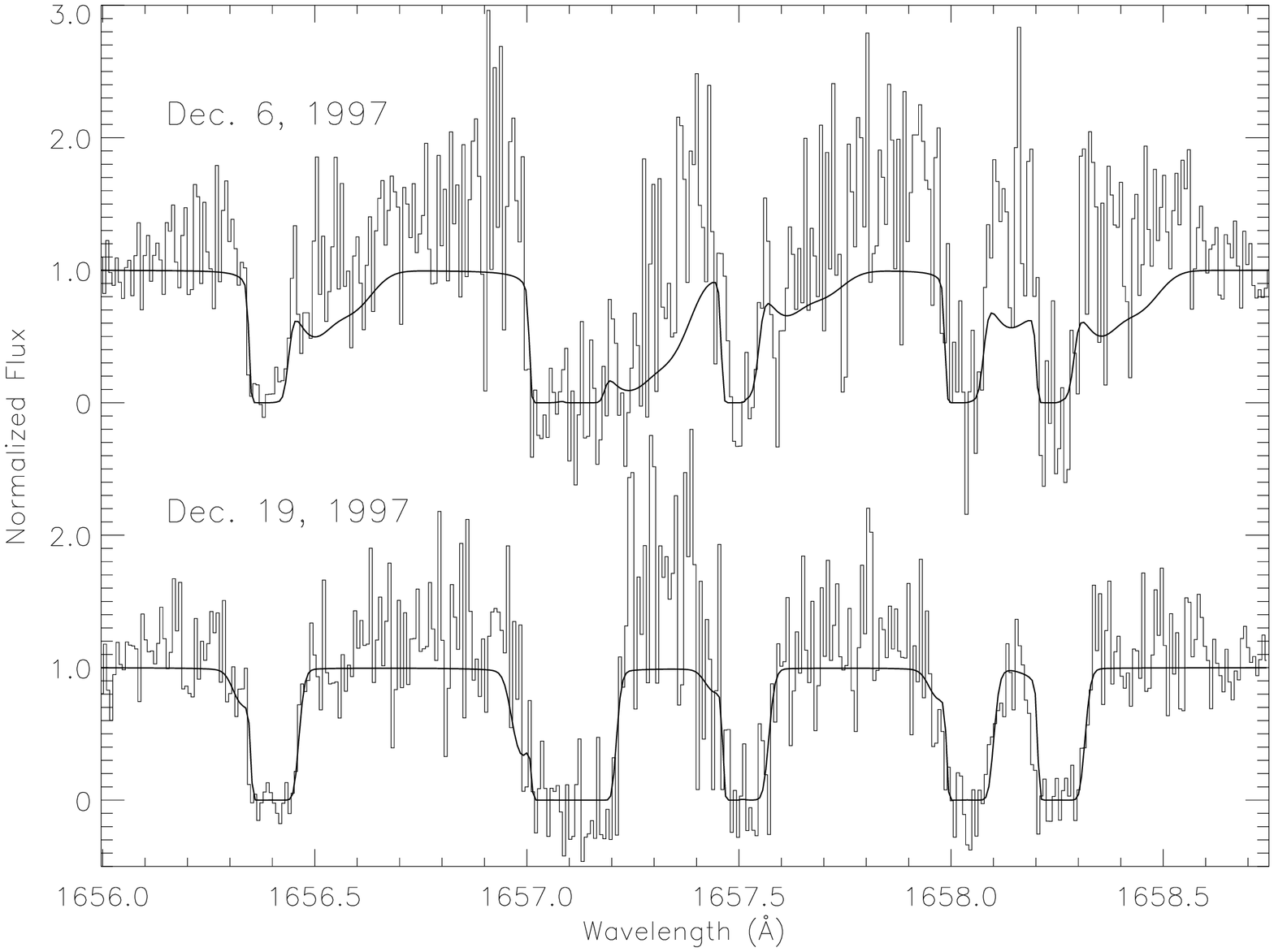, height=5.5in, angle=90}
\caption{The \CI\ $\lambda$1657 multiplet, with both days of 
data shown. The data have not been rebinned or smoothed. The continua around the 
features were fit with sixth-degree polynomials and the best fitting models based 
on the \CIP\ parameters in Table 2 are shown overplotted on the data. \label{fig3}}
\end{center}
\end{figure}

\clearpage
\newpage

\begin{figure}
\begin{center}
\epsfig{file=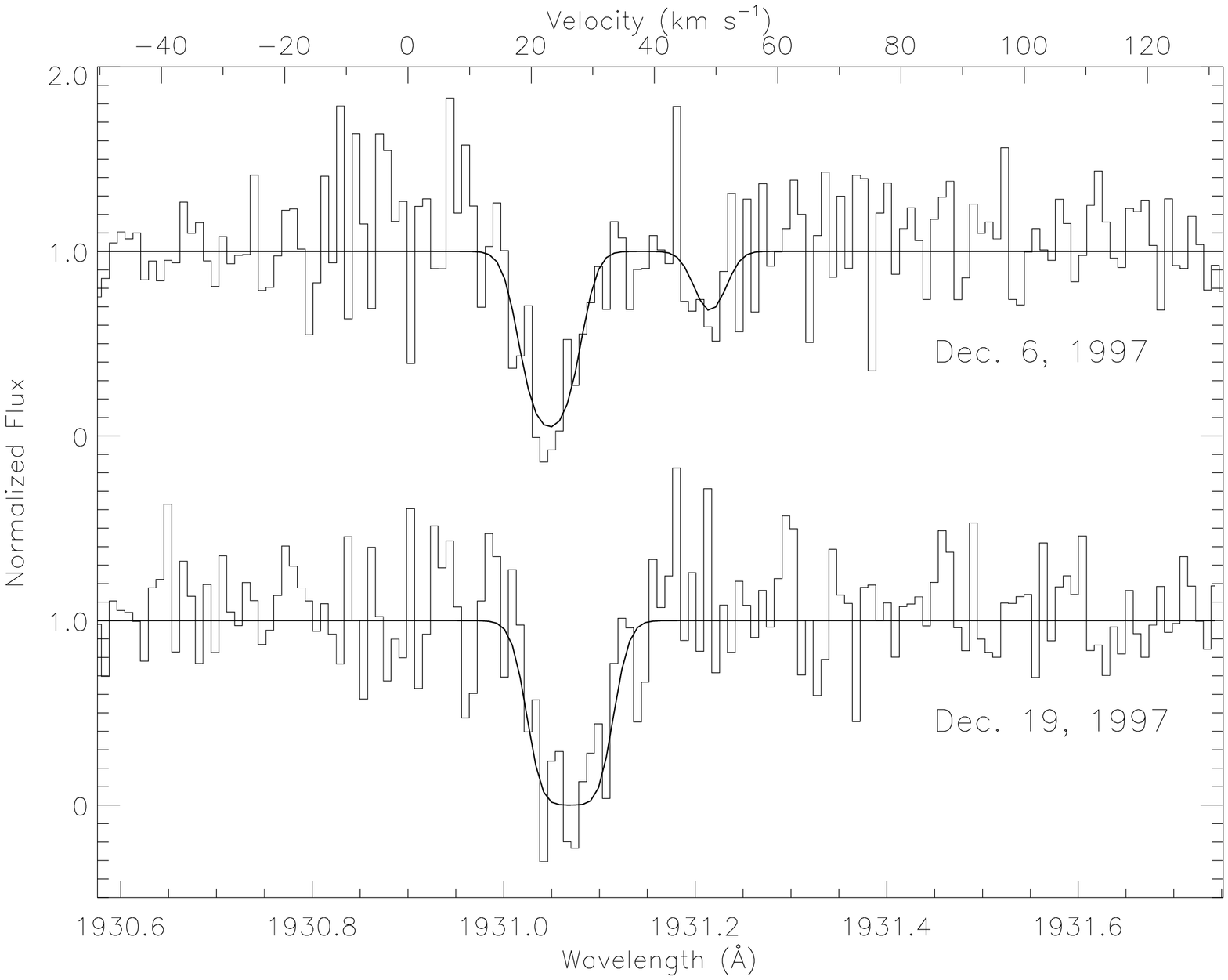, height=5.5in, angle=90}
\caption[figure4.eps]{The \CI\ $\lambda$1931 line, with both days of data 
shown. The data have not been rebinned or smoothed. Heliocentric velocity 
appears on the upper x-axis. The continua around the features were fit with 
fourth-degree polynomials and the best fitting models based on the \CID\ parameters in 
Table 2 are shown overplotted on the data. \label{fig4}}
\end{center}
\end{figure}

\clearpage
\newpage

\begin{figure}
\begin{center}
\epsfig{file=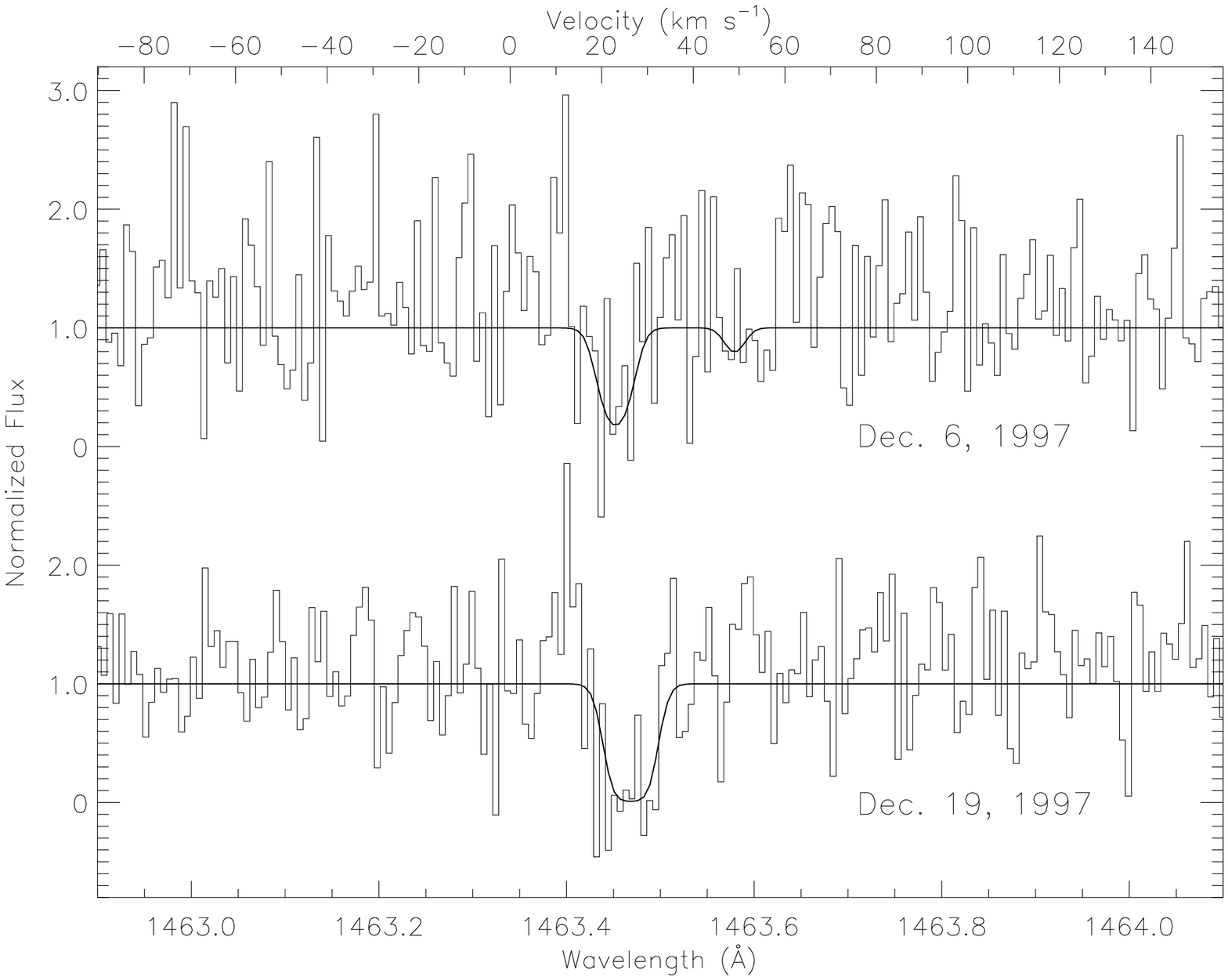, height=5.5in, angle=90}
\caption{The \CI\ $\lambda$1463 line, with both days of data 
shown. The data have not been rebinned or smoothed. Heliocentric velocity 
appears 
on the upper x-axis. The continua around the features were fit with 
fourth-degree 
polynomials and the best fitting models based on the \CID\ parameters in 
Table 2 are shown overplotted on the data. \label{fig5}}
\end{center}
\end{figure}

\clearpage
\newpage

\begin{figure}
\begin{center}
\epsfig{file=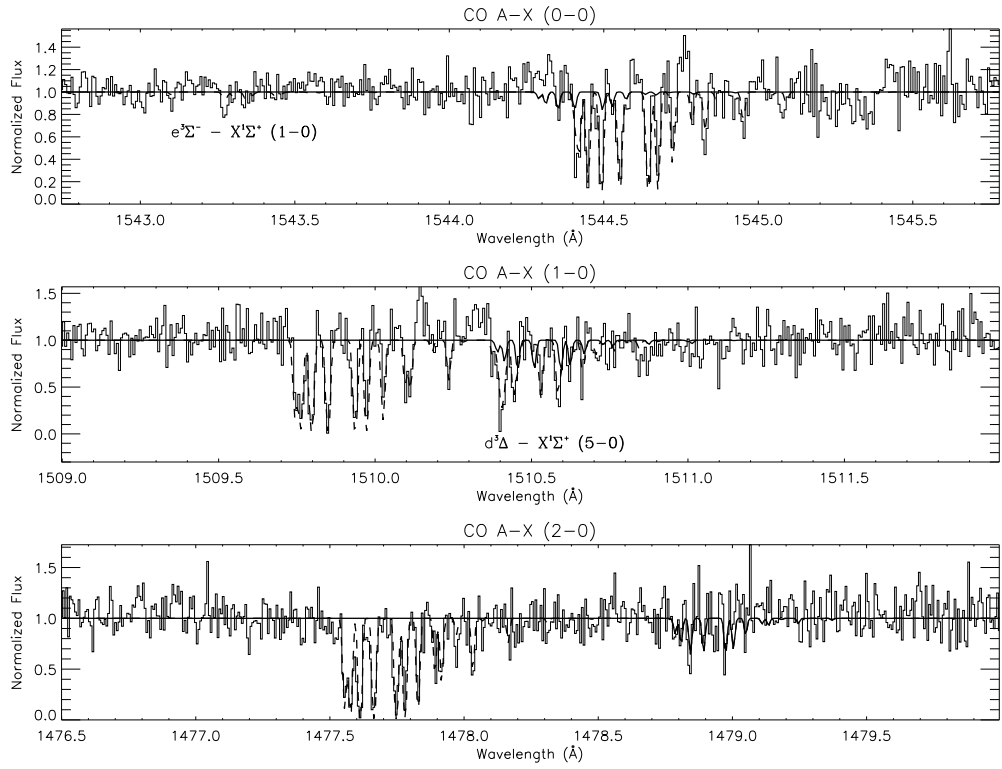, height=5.5in, angle=90}
\caption{The CO $A-X$ (0-0), (1-0), and (2-0) bands, with both 
days of data averaged together. The data have not been rebinned or smoothed. The 
total model based on the CO parameters in Table 2 is overplotted as a dashed line; the 
contribution from $^{13}$CO alone is overplotted as a solid line. \label{fig6}}
\end{center}
\end{figure}

\clearpage
\newpage

\end{document}